\documentclass[preprint2]{aastex}
\slugcomment{Accepted by Astrophysical Journal Letters}
\lefthead{El-Zant et al.}
\righthead{}
\def\fileversion{v1.20a}% was \def\fileversion{v1.20}%
\def\filedate{21.6.94}%  was \def\filedate{26.1.94}%
\edef\epsfigRestoreAt{\catcode`@=\number\catcode`@\relax}%
\catcode`\@=11\relax
\ifx\undefined\@makeother                % -pks-
\def\@makeother#1{\catcode`#1=12\relax}  % -pks-
\fi                                      % -pks-
\immediate\write16{Document style option `epsfig', \fileversion\space
<\filedate> (edited by SPQR + pks)}% was <\filedate> (edited by SPQR)}%
\newcount\EPS@Height \newcount\EPS@Width \newcount\EPS@xscale
\newcount\EPS@yscale
\def\psfigdriver#1{%
  \bgroup\edef\next{\def\noexpand\tempa{#1}}%
    \uppercase\expandafter{\next}%
    \def\LN{DVITOLN03}%
    \def\DVItoPS{DVITOPS}%
    \def\DVIPS{DVIPS}%
    \def\emTeX{EMTEX}%
    \def\OzTeX{OZTEX}%
    \def\Textures{TEXTURES}%
    \global\chardef\fig@driver=0
    \ifx\tempa\LN
        \global\chardef\fig@driver=0\fi
    \ifx\tempa\DVItoPS
        \global\chardef\fig@driver=1\fi
    \ifx\tempa\DVIPS
        \global\chardef\fig@driver=2\fi
    \ifx\tempa\emTeX
        \global\chardef\fig@driver=3\fi
    \ifx\tempa\OzTeX
        \global\chardef\fig@driver=4\fi
    \ifx\tempa\Textures
        \global\chardef\fig@driver=5\fi
  \egroup
\def\psfig@start{}%
\def\psfig@end{}%
\def\epsfig@gofer{}%
\ifcase\fig@driver
\typeout{WARNING! ****
 no specials for LN03 psfig}%
\or % case 1: dvitops
\def\psfig@start{}%
\def\psfig@end{\special{dvitops: import \@p@sfilefinal \space
\@p@swidth sp \space \@p@sheight sp \space fill}%
\if@clip \typeout{Clipping not supported}\fi
\if@angle \typeout{Rotating not supported}\fi
}%
\let\epsfig@gofer\psfig@end
\or %case2 dvips
\def\psfig@start{\special{ps::[begin]  \@p@swidth \space \@p@sheight \space%
        \@p@sbbllx \space \@p@sbblly \space%
        \@p@sbburx \space \@p@sbbury \space%
        startTexFig \space }%
        \if@clip
                \if@verbose
                        \typeout{(clipped to BB) }%
                \fi
                \special{ps:: doclip \space }%
        \fi
        \if@angle              % moved after \if@clip ... \fi -pks-
                \special {ps:: \@p@sangle \space rotate \space}
        \fi
        \special{ps: plotfile \@p@sfilefinal \space }%
        \special{ps::[end] endTexFig \space }%
}%
\def\psfig@end{}%
\def\epsfig@gofer{\if@clip
                        \if@verbose
                           \typeout{(clipped to BB)}%
                        \fi
                        \epsfclipon
                  \fi
                  \epsfsetgraph{\@p@sfilefinal}%
}%
\or % case 3, emTeX
\typeout{WARNING. You must have a .bb info file with the Bounding Box
  of the pcx file}%
\def\psfig@start{}%
\def\psfig@end{\typeout{pcx import of \@p@sfilefinal}%
\if@clip \typeout{Clipping not supported}\fi
\if@angle \typeout{Rotating not supported}\fi
\raisebox{\@p@srheight sp}{\special{em: graph \@p@sfilefinal}}}%
\def\epsfig@gofer{}%
\or % case 4, OzTeX
\def\psfig@start{}%
\def\psfig@end{%
\EPS@Width\@p@swidth
\EPS@Height\@p@sheight
\divide\EPS@Width by 65781  % convert sp to bp
\divide\EPS@Height by 65781
\special{epsf=\@p@sfilefinal
\space
width=\the\EPS@Width
\space
height=\the\EPS@Height
}%
\if@clip \typeout{Clipping not supported}\fi
\if@angle \typeout{Rotating not supported}\fi
}%
\let\epsfig@gofer\psfig@end
\or % case 5, Textures
\def\psfig@end{
         \EPS@Width=\@bbw  
         \divide\EPS@Width by 1000
         \EPS@xscale=\@p@swidth \divide \EPS@xscale by \EPS@Width
         \EPS@Height=\@bbh  
         \divide\EPS@Height by 1000
         \EPS@yscale=\@p@sheight \divide \EPS@yscale by\EPS@Height
  \ifnum\EPS@xscale>\EPS@yscale\EPS@xscale=\EPS@yscale\fi
\if@clip
   \if@verbose
      \typeout{(clipped to BB)}%
   \fi
   \epsfclipon
\fi
\special{illustration \@p@sfilefinal\space scaled \the\EPS@xscale}%
}%
\def\psfig@start{}%
\let\epsfig\psfig
\else
\typeout{WARNING. *** unknown  driver - no psfig}%
\fi
}%
\newdimen\ps@dimcent
\ifx\undefined\fbox
\newdimen\fboxrule
\newdimen\fboxsep
\newdimen\ps@tempdima
\newbox\ps@tempboxa
\long\def\fbox#1{\leavevmode\setbox\ps@tempboxa\hbox{#1}\ps@tempdima\fboxrule
    \advance\ps@tempdima \fboxsep \advance\ps@tempdima \dp\ps@tempboxa
   \hbox{\lower \ps@tempdima\hbox
  {\vbox{\hrule height \fboxrule
          \hbox{\vrule width \fboxrule \hskip\fboxsep
          \vbox{\vskip\fboxsep \box\ps@tempboxa\vskip\fboxsep}\hskip
                 \fboxsep\vrule width \fboxrule}%
                 \hrule height \fboxrule}}}}%
\fi
\ifx\@ifundefined\undefined
\long\def\@ifundefined#1#2#3{\expandafter\ifx\csname
  #1\endcsname\relax#2\else#3\fi}%
\fi
\@ifundefined{typeout}%
{\gdef\typeout#1{\immediate\write\sixt@@n{#1}}}%
{\relax}%
\@ifundefined{epsfig}{}{\typeout{EPSFIG --- already loaded} }%
\@ifundefined{epsfbox}{\input epsf}{}%
\ifx\undefined\@latexerr
        \newlinechar`\^^J
        \def\@spaces{\space\space\space\space}%
        \def\@latexerr#1#2{%
        \edef\@tempc{#2}\expandafter\errhelp\expandafter{\@tempc}%
        \typeout{Error. \space see a manual for explanation.^^J
         \space\@spaces\@spaces\@spaces Type \space H <return> \space for
         immediate help.}\errmessage{#1}}%
\fi
\def\@whattodo{You tried to include a PostScript figure which
cannot be found^^JIf you press return to carry on anyway,^^J
The failed name will be printed in place of the figure.^^J
or type X to quit}%
\def\@whattodobb{You tried to include a PostScript figure which
has no^^Jbounding box, and you supplied none.^^J
If you press return to carry on anyway,^^J
The failed name will be printed in place of the figure.^^J
or type X to quit}%
\def\@nnil{\@nil}%
\def\@empty{}%
\def\@psdonoop#1\@@#2#3{}%
\def\@psdo#1:=#2\do#3{\edef\@psdotmp{#2}\ifx\@psdotmp\@empty \else
    \expandafter\@psdoloop#2,\@nil,\@nil\@@#1{#3}\fi}%
\def\@psdoloop#1,#2,#3\@@#4#5{\def#4{#1}\ifx #4\@nnil \else
       #5\def#4{#2}\ifx #4\@nnil \else#5\@ipsdoloop #3\@@#4{#5}\fi\fi}%
\def\@ipsdoloop#1,#2\@@#3#4{\def#3{#1}\ifx #3\@nnil
       \let\@nextwhile=\@psdonoop \else
      #4\relax\let\@nextwhile=\@ipsdoloop\fi\@nextwhile#2\@@#3{#4}}%
\def\@tpsdo#1:=#2\do#3{\xdef\@psdotmp{#2}\ifx\@psdotmp\@empty \else
    \@tpsdoloop#2\@nil\@nil\@@#1{#3}\fi}%
\def\@tpsdoloop#1#2\@@#3#4{\def#3{#1}\ifx #3\@nnil
       \let\@nextwhile=\@psdonoop \else
      #4\relax\let\@nextwhile=\@tpsdoloop\fi\@nextwhile#2\@@#3{#4}}%
\long\def\epsfaux#1#2:#3\\{\ifx#1\epsfpercent
   \def\testit{#2}\ifx\testit\epsfbblit
        \@atendfalse
        \epsf@atend #3 . \\%
        \if@atend
           \if@verbose
                \typeout{epsfig: found `(atend)'; continuing search}%
           \fi
        \else
                \epsfgrab #3 . . . \\%
                \epsffileokfalse\global\no@bbfalse
                \global\epsfbbfoundtrue
        \fi
   \fi\fi}%
\def\epsf@atendlit{(atend)}
\def\epsf@atend #1 #2 #3\\{%
   \def\epsf@tmp{#1}\ifx\epsf@tmp\empty
      \epsf@atend #2 #3 .\\\else
   \ifx\epsf@tmp\epsf@atendlit\@atendtrue\fi\fi}%

\chardef\trig@letter = 11
\chardef\other = 12
 
\newif\ifdebug %%% turn me on to see TeX hard at work ...
\newif\ifc@mpute %%% don't need to compute some values
\newif\if@atend
\c@mputetrue % but assume that we do
 
\let\then = \relax
\def\r@dian{pt }%
\let\r@dians = \r@dian
\let\dimensionless@nit = \r@dian
\let\dimensionless@nits = \dimensionless@nit
\def\internal@nit{sp }%
\let\internal@nits = \internal@nit
\newif\ifstillc@nverging
\def \Mess@ge #1{\ifdebug \then \message {#1} \fi}%
 
{ %%% Things that need abnormal catcodes %%%
        \catcode `\@ = \trig@letter
        \gdef \nodimen {\expandafter \n@dimen \the \dimen}%
        \gdef \term #1 #2 #3%
               {\edef \t@ {\the #1}%%% freeze parameter 1 (count, by value)
                \edef \t@@ {\expandafter \n@dimen \the #2\r@dian}%
                \t@rm {\t@} {\t@@} {#3}%
               }%
        \gdef \t@rm #1 #2 #3%
               {{%
                \count 0 = 0
                \dimen 0 = 1 \dimensionless@nit
                \dimen 2 = #2\relax
                \Mess@ge {Calculating term #1 of \nodimen 2}%
                \loop
                \ifnum  \count 0 < #1
                \then   \advance \count 0 by 1
                        \Mess@ge {Iteration \the \count 0 \space}%
                        \Multiply \dimen 0 by {\dimen 2}%
                        \Mess@ge {After multiplication, term = \nodimen 0}%
                        \Divide \dimen 0 by {\count 0}%
                        \Mess@ge {After division, term = \nodimen 0}%
                \repeat
                \Mess@ge {Final value for term #1 of
                                \nodimen 2 \space is \nodimen 0}%
                \xdef \Term {#3 = \nodimen 0 \r@dians}%
                \aftergroup \Term
               }}%
        \catcode `\p = \other
        \catcode `\t = \other
        \gdef \n@dimen #1pt{#1} %%% throw away the ``pt''
}%
 
\def \Divide #1by #2{\divide #1 by #2} %%% just a synonym
 
\def \Multiply #1by #2%%% allows division of a dimen by a dimen
       {{%%% should really freeze parameter 2 (dimen, passed by value)
        \count 0 = #1\relax
        \count 2 = #2\relax
        \count 4 = 65536
        \Mess@ge {Before scaling, count 0 = \the \count 0 \space and
                        count 2 = \the \count 2}%
        \ifnum  \count 0 > 32767 %%% do our best to avoid overflow
        \then   \divide \count 0 by 4
                \divide \count 4 by 4
        \else   \ifnum  \count 0 < -32767
                \then   \divide \count 0 by 4
                        \divide \count 4 by 4
                \else
                \fi
        \fi
        \ifnum  \count 2 > 32767 %%% while retaining reasonable accuracy
        \then   \divide \count 2 by 4
                \divide \count 4 by 4
        \else   \ifnum  \count 2 < -32767
                \then   \divide \count 2 by 4
                        \divide \count 4 by 4
                \else
                \fi
        \fi
        \multiply \count 0 by \count 2
        \divide \count 0 by \count 4
        \xdef \product {#1 = \the \count 0 \internal@nits}%
        \aftergroup \product
       }}%
 
\def\r@duce{\ifdim\dimen0 > 90\r@dian \then   % sin(x) = sin(180-x)
                \multiply\dimen0 by -1
                \advance\dimen0 by 180\r@dian
                \r@duce
            \else \ifdim\dimen0 < -90\r@dian \then  % sin(x) = sin(360+x)
                \advance\dimen0 by 360\r@dian
                \r@duce
                \fi
            \fi}%
 
\def\Sine#1%
       {{%
        \dimen 0 = #1 \r@dian
        \r@duce
        \ifdim\dimen0 = -90\r@dian \then
           \dimen4 = -1\r@dian
           \c@mputefalse
        \fi
        \ifdim\dimen0 = 90\r@dian \then
           \dimen4 = 1\r@dian
           \c@mputefalse
        \fi
        \ifdim\dimen0 = 0\r@dian \then
           \dimen4 = 0\r@dian
           \c@mputefalse
        \fi
        \ifc@mpute \then
                \divide\dimen0 by 180
                \dimen0=3.141592654\dimen0
                \dimen 2 = 3.1415926535897963\r@dian %%% a well-known constant
                \divide\dimen 2 by 2 %%% we only deal with -pi/2 : pi/2
                \Mess@ge {Sin: calculating Sin of \nodimen 0}%
                \count 0 = 1 %%% see power-series expansion for sine
                \dimen 2 = 1 \r@dian %%% ditto
                \dimen 4 = 0 \r@dian %%% ditto
                \loop
                        \ifnum  \dimen 2 = 0 %%% then we've done
                        \then   \stillc@nvergingfalse
                        \else   \stillc@nvergingtrue
                        \fi
                        \ifstillc@nverging %%% then calculate next term
                        \then   \term {\count 0} {\dimen 0} {\dimen 2}%
                                \advance \count 0 by 2
                                \count 2 = \count 0
                                \divide \count 2 by 2
                                \ifodd  \count 2 %%% signs alternate
                                \then   \advance \dimen 4 by \dimen 2
                                \else   \advance \dimen 4 by -\dimen 2
                                \fi
                \repeat
        \fi
                        \xdef \sine {\nodimen 4}%
       }}%
 
\def\Cosine#1{\ifx\sine\UnDefined\edef\Savesine{\relax}\else
                             \edef\Savesine{\sine}\fi
        {\dimen0=#1\r@dian\multiply\dimen0 by -1
         \advance\dimen0 by 90\r@dian
         \Sine{\nodimen 0}%
         \xdef\cosine{\sine}%
         \xdef\sine{\Savesine}}}
\def\psdraft{\def\@psdraft{0}}%
\def\psfull{\def\@psdraft{1}}%
\psfull
\newif\if@compress
\def\pscompress{\@compresstrue}
\def\psnocompress{\@compressfalse}
\@compressfalse
\newif\if@scalefirst
\def\psscalefirst{\@scalefirsttrue}%
\def\psrotatefirst{\@scalefirstfalse}%
\psrotatefirst
\newif\if@draftbox
\def\psnodraftbox{\@draftboxfalse}%
\@draftboxtrue
\newif\if@noisy
\@noisyfalse
\newif\ifno@bb
\newif\if@bbllx
\newif\if@bblly
\newif\if@bburx
\newif\if@bbury
\newif\if@height
\newif\if@width
\newif\if@rheight
\newif\if@rwidth
\newif\if@angle
\newif\if@clip
\newif\if@verbose
\newif\if@prologfile
\def\@p@@sprolog#1{\@prologfiletrue\def\@prologfileval{#1}}%
\def\@p@@sclip#1{\@cliptrue}%
\newif\ifepsfig@dos  % only single suffix possible
\def\epsfigdos{\epsfig@dostrue}%
\epsfig@dosfalse
\newif\ifuse@psfig
\def\ParseName#1{\expandafter\@Parse#1}%
\def\@Parse#1.#2:{\gdef\BaseName{#1}\gdef\FileType{#2}}%

\def\@p@@sfile#1{%
  \ifepsfig@dos
     \ParseName{#1:}%
  \else
     \gdef\BaseName{#1}\gdef\FileType{}%
  \fi
  \def\@p@sfile{NO FILE: #1}%
  \def\@p@sfilefinal{NO FILE: #1}%
  \openin1=#1
  \ifeof1\closein1\openin1=\BaseName.bb
    \ifeof1\closein1
      \if@bbllx                 % No postscript file but bb given explicitly.
        \if@bblly\if@bburx\if@bbury
          \def\@p@sfile{#1}%
          \def\@p@sfilefinal{#1}%
        \fi\fi\fi
      \else                     % No bounding box found.
        \@latexerr{ERROR. PostScript file #1 not found}\@whattodo
        \@p@@sbbllx{100bp}%
        \@p@@sbblly{100bp}%
        \@p@@sbburx{200bp}%
        \@p@@sbbury{200bp}%
        \psdraft
      \fi
    \else                       % Postscript file is compressed.
      \closein1%
      \edef\@p@sfile{\BaseName.bb}%
      \typeout{using BB from \@p@sfile}%
      \ifnum\fig@driver=3
        \edef\@p@sfilefinal{\BaseName.pcx}%
      \else
        \ifepsfig@dos
          \edef\@p@sfilefinal{"`gunzip -c `texfind \BaseName.{z,Z,gz}"}%
        \else
          \edef\@p@sfilefinal{"`epsfig \if@compress-c \fi#1"}%          
        \fi
      \fi
    \fi
  \else\closein1                % Postscript file is not compressed.
    \edef\@p@sfile{#1}%
    \if@compress  
      \edef\@p@sfilefinal{"`epsfig -c #1"}%
    \else
      \edef\@p@sfilefinal{#1}%
    \fi
  \fi%
}

\let\@p@@sfigure\@p@@sfile
\def\@p@@sbbllx#1{%
                                            \@bbllxtrue
                \ps@dimcent=#1
                \edef\@p@sbbllx{\number\ps@dimcent}%
                \divide\ps@dimcent by65536
                \global\edef\epsfllx{\number\ps@dimcent}%
}%
\def\@p@@sbblly#1{%
                \@bbllytrue
                \ps@dimcent=#1
                \edef\@p@sbblly{\number\ps@dimcent}%
                \divide\ps@dimcent by65536
                \global\edef\epsflly{\number\ps@dimcent}%
}%
\def\@p@@sbburx#1{%
                \@bburxtrue
                \ps@dimcent=#1
                \edef\@p@sbburx{\number\ps@dimcent}%
                \divide\ps@dimcent by65536
                \global\edef\epsfurx{\number\ps@dimcent}%
}%
\def\@p@@sbbury#1{%
                \@bburytrue
                \ps@dimcent=#1
                \edef\@p@sbbury{\number\ps@dimcent}%
                \divide\ps@dimcent by65536
                \global\edef\epsfury{\number\ps@dimcent}%
}%
\def\@p@@sheight#1{%
                \@heighttrue
                \global\epsfysize=#1
                \ps@dimcent=#1
                \edef\@p@sheight{\number\ps@dimcent}%
}%
\def\@p@@swidth#1{%
                \@widthtrue
                \global\epsfxsize=#1
                \ps@dimcent=#1
                \edef\@p@swidth{\number\ps@dimcent}% 
}%
\def\@p@@srheight#1{%
                \@rheighttrue\use@psfigtrue
                \ps@dimcent=#1
                \edef\@p@srheight{\number\ps@dimcent}%
}%
\def\@p@@srwidth#1{%
                \@rwidthtrue\use@psfigtrue
                \ps@dimcent=#1
                \edef\@p@srwidth{\number\ps@dimcent}%
}%
\def\@p@@sangle#1{%
                \use@psfigtrue
                \@angletrue
                \edef\@p@sangle{#1}%
}%
\def\@p@@ssilent#1{%
                \@verbosefalse
}%
\def\@p@@snoisy#1{%
                \@verbosetrue
}%
\def\@cs@name#1{\csname #1\endcsname}%
\def\@setparms#1=#2,{\@cs@name{@p@@s#1}{#2}}%
\def\ps@init@parms{%
                \@bbllxfalse \@bbllyfalse
                \@bburxfalse \@bburyfalse
                \@heightfalse \@widthfalse
                \@rheightfalse \@rwidthfalse
                \def\@p@sbbllx{}\def\@p@sbblly{}%
                \def\@p@sbburx{}\def\@p@sbbury{}%
                \def\@p@sheight{}\def\@p@swidth{}%
                \def\@p@srheight{}\def\@p@srwidth{}%
                \def\@p@sangle{0}%
                \def\@p@sfile{}%
                \use@psfigfalse
                \@prologfilefalse
                \def\@sc{}%
                \if@noisy
                        \@verbosetrue
                \else
                        \@verbosefalse
                \fi
                \@clipfalse
}%
\def\parse@ps@parms#1{%
                \@psdo\@psfiga:=#1\do
                   {\expandafter\@setparms\@psfiga,}%
\if@prologfile
\fi
}%
\def\bb@missing{%
        \if@verbose
            \typeout{psfig: searching \@p@sfile \space  for bounding box}%
        \fi
        \epsfgetbb{\@p@sfile}%
        \ifepsfbbfound
            \ps@dimcent=\epsfllx bp\edef\@p@sbbllx{\number\ps@dimcent}%
            \ps@dimcent=\epsflly bp\edef\@p@sbblly{\number\ps@dimcent}%
            \ps@dimcent=\epsfurx bp\edef\@p@sbburx{\number\ps@dimcent}%
            \ps@dimcent=\epsfury bp\edef\@p@sbbury{\number\ps@dimcent}%
        \else
            \epsfbbfoundfalse
        \fi
}
\newdimen\p@intvaluex
\newdimen\p@intvaluey
\def\rotate@#1#2{{\dimen0=#1 sp\dimen1=#2 sp
                  \global\p@intvaluex=\cosine\dimen0
                  \dimen3=\sine\dimen1
                  \global\advance\p@intvaluex by -\dimen3
                  \global\p@intvaluey=\sine\dimen0
                  \dimen3=\cosine\dimen1
                  \global\advance\p@intvaluey by \dimen3
                  }}%
\def\compute@bb{%
                \epsfbbfoundfalse
                \if@bbllx\epsfbbfoundtrue\fi
                \if@bblly\epsfbbfoundtrue\fi
                \if@bburx\epsfbbfoundtrue\fi
                \if@bbury\epsfbbfoundtrue\fi
                \ifepsfbbfound\else\bb@missing\fi
                \ifepsfbbfound\else
                \@latexerr{ERROR. cannot locate BoundingBox}\@whattodobb
                        \@p@@sbbllx{100bp}%
                        \@p@@sbblly{100bp}%
                        \@p@@sbburx{200bp}%
                        \@p@@sbbury{200bp}%
                        \no@bbtrue
                        \psdraft
                \fi
                \count203=\@p@sbburx
                \count204=\@p@sbbury
                \advance\count203 by -\@p@sbbllx
                \advance\count204 by -\@p@sbblly
                \edef\ps@bbw{\number\count203}%
                \edef\ps@bbh{\number\count204}%
                 \edef\@bbw{\number\count203}%
                \edef\@bbh{\number\count204}%
               \if@angle
                        \Sine{\@p@sangle}\Cosine{\@p@sangle}%
 
{\ps@dimcent=\maxdimen\xdef\r@p@sbbllx{\number\ps@dimcent}%
 
\xdef\r@p@sbblly{\number\ps@dimcent}%
 
\xdef\r@p@sbburx{-\number\ps@dimcent}%
 
\xdef\r@p@sbbury{-\number\ps@dimcent}}%
                        \def\minmaxtest{%
                           \ifnum\number\p@intvaluex<\r@p@sbbllx
                              \xdef\r@p@sbbllx{\number\p@intvaluex}\fi
                           \ifnum\number\p@intvaluex>\r@p@sbburx
                              \xdef\r@p@sbburx{\number\p@intvaluex}\fi
                           \ifnum\number\p@intvaluey<\r@p@sbblly
                              \xdef\r@p@sbblly{\number\p@intvaluey}\fi
                           \ifnum\number\p@intvaluey>\r@p@sbbury
                              \xdef\r@p@sbbury{\number\p@intvaluey}\fi
                           }%
                        \rotate@{\@p@sbbllx}{\@p@sbblly}%
                        \minmaxtest
                        \rotate@{\@p@sbbllx}{\@p@sbbury}%
                        \minmaxtest
                        \rotate@{\@p@sbburx}{\@p@sbblly}%
                        \minmaxtest
                        \rotate@{\@p@sbburx}{\@p@sbbury}%
                        \minmaxtest
 
\edef\@p@sbbllx{\r@p@sbbllx}\edef\@p@sbblly{\r@p@sbblly}%
 
\edef\@p@sbburx{\r@p@sbburx}\edef\@p@sbbury{\r@p@sbbury}%
                \fi
                \count203=\@p@sbburx
                \count204=\@p@sbbury
                \advance\count203 by -\@p@sbbllx
                \advance\count204 by -\@p@sbblly
                \edef\@bbw{\number\count203}%
                \edef\@bbh{\number\count204}%
}%
\def\in@hundreds#1#2#3{\count240=#2 \count241=#3
                     \count100=\count240        % 100 is first digit #2/#3
                     \divide\count100 by \count241
                     \count101=\count100
                     \multiply\count101 by \count241
                     \advance\count240 by -\count101
                     \multiply\count240 by 10
                     \count101=\count240        %101 is second digit of #2/#3
                     \divide\count101 by \count241
                     \count102=\count101
                     \multiply\count102 by \count241
                     \advance\count240 by -\count102
                     \multiply\count240 by 10
                     \count102=\count240        % 102 is the third digit
                     \divide\count102 by \count241
                     \count200=#1\count205=0
                     \count201=\count200
                        \multiply\count201 by \count100
                        \advance\count205 by \count201
                     \count201=\count200
                        \divide\count201 by 10
                        \multiply\count201 by \count101
                        \advance\count205 by \count201
                     \count201=\count200
                        \divide\count201 by 100
                        \multiply\count201 by \count102
                        \advance\count205 by \count201
                     \edef\@result{\number\count205}%
}%
\def\compute@wfromh{%
                \in@hundreds{\@p@sheight}{\@bbw}{\@bbh}%
                \edef\@p@swidth{\@result}%
}%
\def\compute@hfromw{%
                \in@hundreds{\@p@swidth}{\@bbh}{\@bbw}%
                \edef\@p@sheight{\@result}%
}%
\def\compute@handw{%
                \if@height
                        \if@width
                        \else
                                \compute@wfromh
                        \fi
                \else
                        \if@width
                                \compute@hfromw
                        \else
                                \edef\@p@sheight{\@bbh}%
                                \edef\@p@swidth{\@bbw}%
                        \fi
                \fi
}%
\def\compute@resv{%
                \if@rheight \else \edef\@p@srheight{\@p@sheight} \fi
                \if@rwidth \else \edef\@p@srwidth{\@p@swidth} \fi
}%
\def\compute@sizes{%
        \if@scalefirst\if@angle
        \if@width
           \in@hundreds{\@p@swidth}{\@bbw}{\ps@bbw}%
           \edef\@p@swidth{\@result}%
        \fi
        \if@height
           \in@hundreds{\@p@sheight}{\@bbh}{\ps@bbh}%
           \edef\@p@sheight{\@result}%
        \fi
        \fi\fi
        \compute@handw
        \compute@resv
}
\long\def\graphic@verb#1{\def\next{#1}%
  {\expandafter\graphic@strip\meaning\next}}
\def\graphic@strip#1>{}
\def\graphic@zapspace#1{%
  #1\ifx\graphic@zapspace#1\graphic@zapspace%
  \else\expandafter\graphic@zapspace%
  \fi}
\def\psfig#1{%
\edef\@tempa{\graphic@zapspace#1{}}%
\ifvmode\leavevmode\fi\vbox {%
        \ps@init@parms
        \parse@ps@parms{\@tempa}%
        \ifnum\@psdraft=1
                \typeout{[\@p@sfilefinal]}%
                \if@verbose
                        \typeout{epsfig: using PSFIG macros}%
                \fi
                \psfig@method
        \else
                \epsfig@draft
        \fi
}
}%
\def\graphic@zapspace#1{%
  #1\ifx\graphic@zapspace#1\graphic@zapspace%
  \else\expandafter\graphic@zapspace%
  \fi}
\def\epsfig#1{%
\edef\@tempa{\graphic@zapspace#1{}}%
\ifvmode\leavevmode\fi\vbox {%
        \ps@init@parms
        \parse@ps@parms{\@tempa}%
        \ifnum\@psdraft=1
          \if@angle\use@psfigtrue\fi
          {\ifnum\fig@driver=1\global\use@psfigtrue\fi}%
          {\ifnum\fig@driver=3\global\use@psfigtrue\fi}%
          {\ifnum\fig@driver=4\global\use@psfigtrue\fi}%
          {\ifnum\fig@driver=5\global\use@psfigtrue\fi}%
                \ifuse@psfig
                        \if@verbose
                                \typeout{epsfig: using PSFIG macros}%
                        \fi
                        \psfig@method
                \else
                        \if@verbose
                                \typeout{epsfig: using EPSF macros}%
                        \fi
                        \epsf@method
                \fi
        \else
                \epsfig@draft
        \fi
}%
}%

\def\epsf@method{%
        \epsfbbfoundfalse
        \if@bbllx\epsfbbfoundtrue\fi
        \if@bblly\epsfbbfoundtrue\fi
        \if@bburx\epsfbbfoundtrue\fi
        \if@bbury\epsfbbfoundtrue\fi
        \ifepsfbbfound\else\epsfgetbb{\@p@sfile}\fi
        \ifepsfbbfound
           \typeout{<\@p@sfilefinal>}%
           \epsfig@gofer
        \else
          \@latexerr{ERROR - Cannot locate BoundingBox}\@whattodobb
          \@p@@sbbllx{100bp}%
          \@p@@sbblly{100bp}%
          \@p@@sbburx{200bp}%
          \@p@@sbbury{200bp}%
                \count203=\@p@sbburx
                \count204=\@p@sbbury
                \advance\count203 by -\@p@sbbllx
                \advance\count204 by -\@p@sbblly
                \edef\@bbw{\number\count203}%
                \edef\@bbh{\number\count204}%
          \compute@sizes
          \epsfig@@draft
       \fi
}%
\def\psfig@method{%
        \compute@bb
        \ifepsfbbfound
          \compute@sizes
          \psfig@start
          \vbox to \@p@srheight sp{\hbox to \@p@srwidth 
            sp{\hss}\vss\psfig@end}%
        \else
           \epsfig@draft
        \fi
}%
\def\epsfig@draft{\compute@bb\compute@sizes\epsfig@@draft}%
\def\epsfig@@draft{%
\typeout{<(draft only) \@p@sfilefinal>}%
\if@draftbox
        \hbox{{\fboxsep0pt\fbox{\vbox to \@p@srheight sp{%
        \vss\hbox to \@p@srwidth sp{ \hss 
           \expandafter\Literally\@p@sfilefinal\@nil
                          \hss }\vss
        }}}}%
\else
        \vbox to \@p@srheight sp{%
        \vss\hbox to \@p@srwidth sp{\hss}\vss}%
\fi
}%
\def\Literally#1\@nil{{\tt\graphic@verb{#1}}}
\psfigdriver{dvips}%
\epsfigRestoreAt

\def\kmsmpc{\mathrm{km}\ \mathrm{s^{-1}} \mathrm{Mpc^{-1}}}

\def\hkpc{h^{-1} \mathrm{kpc}}

\def\hmsun{h^{-1} M_\odot}
 
\def \m3{{\rm Mark III}}
\def \etal {{\it et al.\ }}

\begin{document}
\lefthead{El-Zant et al.}
\righthead{}

\title{Flat-Cored Dark Matter in Cuspy Clusters of Galaxies}

\author{
Amr A. El-Zant,\altaffilmark{1} Yehuda Hoffman,\altaffilmark{2}
Joel Primack,\altaffilmark{3}  Francoise Combes,\altaffilmark{4}
Isaac Shlosman\altaffilmark{5}
}

\altaffiltext{1}{
Theoretical Astrophysics 130-33, Caltech, Pasadena, CA
91125, USA}
\altaffiltext{2}{Racah Institute of Physics, Hebrew University,
Jerusalem 91904, Israel}
\altaffiltext{3}{Physics Department, University of California, Santa
Cruz, CA 95064, USA}
\altaffiltext{4}{LERMA, Observatoire de Paris, Paris, France}
\altaffiltext{5}{Department of Physics and Astronomy,
University of Kentucky, Lexington, KY 40506-0055, USA}

\begin{abstract}
Sand, Treu \& Ellis (2002) have measured the central density profile
of cluster MS2137-23 with gravitational lensing and velocity
dispersion and removed the stellar contribution with a reasonable
$M/L$. The resulting dark matter (DM)  distribution within $r<50\hkpc$ was
fitted by a density cusp of $r^{-\beta}$ with $\beta=0.35$, in an
apparent contradiction to the CDM prediction of $\beta\sim1$. The
disagreement worsens if adiabatic compression of the DM by the infalling
baryons is considered. Following El-Zant, Shlosman \& Hoffman (2001),
we argue that dynamical friction acting on galaxies
moving within the DM background counters the effect of adiabatic
compression by transfering their orbital energy to the DM, thus
heating up and softening the cusp. Using
$N$-body simulations we show that indeed the inner DM distribution flattens
(with $\beta\approx 0.35$ for a cluster like MS2137-23) when the galaxies
spiral inward. We find as a robust result that while the DM distribution
becomes core-like, the overall mass distribution preserves its
cuspy nature, in agreement with X-ray and lensing observations of clusters.
\end{abstract}

%%%%%%%%%%%%%%%%%%%%%%%%%%%%%%%%%%

\section{Introduction}
\label{sec:intro}

The  cold dark matter ($\Lambda$CDM) cosmology (Blumenthal \etal
1984; Davis \etal 1985) with a large cosmological constant
$\Omega_\Lambda \approx 0.7$ is in excellent
agreement with data on large-scale structure and cosmic microwave
background anisotropies (Spergel \etal 2003). Yet, concerns have been raised
about the possible
disagreement between numerical simulations of DM and
observations of the
centers of galaxies and clusters (Flores \& Primack 1994; Moore 1994).
CDM, and CDM-like, simulations all produce DM halos characterized by a
divergent, ``cuspy,'' inner density profile (Navarro, Frenk \& White
1997; NFW), while observations seem to indicate that at least some
galaxies and clusters have flat, core-like, inner density profiles
(e.g., Primack 2003). Poorly understood galaxy formation
processes and complicated galactic dynamics make the comparison of
simulations and observations of galaxies more difficult to
interpret. In contrast, clusters provide a much `cleaner' environment
to test the CDM model. In particular, observational determination of
the mass distribution in clusters by gravitational lensing provides
robust results and confrontation of these with simulations can allow
strong tests of the CDM cosmogony.

Sand, Treu \& Ellis (2002; STE) measured the density profile in the
central $50 \hkpc$ of cluster MS2137-23 at $z=0.313$ using
gravitationally lensed radial and tangential arcs to determine the
central radial profile and total mass enclosed respectively, and the
velocity dispersion profile of the central galaxy from a long-slit
spectrum as an additional constraint.  ($h$ is the
Hubble constant in units of $100\ \kmsmpc$.)    The stellar contribution to the
mass was fitted with a
reasonable $M_*/L_V = 3.1 M_\odot/L_\odot$, which resulted in
$r^{-\beta}$ DM central density profile, with a best fit value $\beta=
0.35$ and a 99\% C.L. upper limit $\beta<0.9$.  This is inconsistent
with dissipationless CDM simulations, which give $\beta =1-1.5$ (Moore
\etal 1999; Power \etal 2003; Navarro et al. 2003).
STE argue that baryonic infall should steepen the DM central
cusp, increasing $\beta$, via adiabatic compression, thus worsening
the disagreement.
Sand \etal (2003a,b) have presented data on five additional clusters,
most of which have inner DM profiles similar to MS2137 --
although the cluster with the lowest luminosity central galaxy, RXJ
1133, has a best fit $\beta \approx 1$. Kelson \etal (2002) used
stellar velocities to trace the central mass distribution of Abell
2199 with similar conclusions. Here we  argue that this apparent
contradiction can be naturally resolved within the framework
of the CDM model once the effects of the baryons are properly taken
into account.

The DM adiabatic compression induced by gaseous baryonic infall was
used by Blumenthal \etal (1986), Flores \etal (1993), and Mo, Mao \&
White (1998) to describe the effects on the DM distribution of baryon
settling in virialized DM halos. Slowly cooling baryons radiate energy
sinking to the bottom, steepening the total density profile.
This scenario was questioned
for galaxies in the context of inhomogeneous baryonic collapse (El-Zant,
Shlosman \& Hoffman 2001; ESH) and in the context of CDM hierarchical halo
formation by mergers (Vitvitska \etal 2002).

Once the baryons condense to form stars and galaxies, they experience
a dynamical friction (DF) from the far less massive DM particles, as they move
through the halo (Chandrasekhar 1943). Early-type galaxies,
being  denser than the DM in cluster halos,  survive tidal interactions.
The energy  dissipated in this case is transferred to
the DM, increasing its random motion, rather than being lost by the
system. Some elements of this effect were considered by Merritt (1983) who
studied tidal stripping of galaxies in clusters in a different context.

Most  studies of the role of DF  in the evolution of groups and
clusters of galaxies have focused on the fate of the
infalling galaxies (e.g., Dubinski 1998).  ESH on the other hand, adopted a
different approach and analyzed the DF-induced changes in the DM halo
structure within the context of galaxy formation,
where gas cools to form dense clumps in a cuspy
NFW halo.  Specifically, they have shown that the
orbital energy lost by the clumps to the DM background is sufficient
to ``heat-up'' and flatten the DM density cusps. This DF heating
dominates over adiabatic contraction.  The semi-analytical Monte-Carlo
approach of  ESH has been confirmed by $N$-body simulations, with massive
baryonic clumps represented by finite-size solid particles (El-Zant \etal
2004). Following ESH, we investigate the role of DF in flattening of cuspy
density profiles in clusters of galaxies by means of $N$-body simulations.

\section{Numerical Modeling}
\label{sec:DF}

The dynamics of a system made of 
$N=9\times 10^5$
DM particles, out of which
$N_{clump}$ are massive clumps of mass $M_{clump}$,  is followed by a PM
(particle mesh) code using a non-periodic 256$^3$
useful grid.  The ratio of the mass in clumps to the total mass
in the simulation is allowed to vary
from the global baryonic-to-total mass ratio of  $0.16$ recently reported
by the WMAP collaboration  (Spergel \etal\ 2003),
hence relaxing the assumption that all the baryons are locked in the
clumps. The force resolution, which implies a gravitational softening,
corresponds to $1/39$ of the cluster's NFW scale length $r_s$. Particles that
get outside the PM region  are advanced with the approximation
of Keplerian motion, with all the mass of other particles assumed
to be gathered in a point mass.

The system is initially in virial equilibrium and the baryonic clump
distribution follows that of the DM, which has the NFW density profile
    \begin{equation}
      \rho(r)=\rho_s r_s^3 /r(r_s+r)^2,
        \label{eq:NFW}
    \end{equation}
where $r_s$ is the radius at which the logarithmic derivative equals $-2$.
The initial isothropic velocity dispersion is determined by solving the Jeans
equations --- a valid assumption in the central region of the cluster, in
agreement with $N$-body simulations (Colin \etal 2000), and also
motions of central cluster galaxies (van der Marel \etal 2000).  The
concentration parameter
is defined as $c\equiv R_{vir}/r_s$, where $R_{vir}$ is the halo virial
radius.  Following ESH, where the DF is
found to be effective in modifying
the halo profile only {\it within} $r <  r_s$, the isotropy assumption
should not affect our results.

The DF is expected to affect the dynamics of the very inner region, well within
$r_s$, making the effect studied here essentially independent of the dynamics
outside of $r_s$. To the extent that the initial NFW configuration remains
numerically  stable, one can apply the scaling properties of the NFW profile
to the numerical simulations. An initial spherical NFW  halo with $R_{vir}=c_0
r_s$  can be rescaled to any virtual NFW model, provided that its
concentration parameter is larger than  $c_0$, by undergoing a suitable
rescaling of the time (or density). This rescaling can be extended to our case
as long as the ratio $M_{clump}$ to the total mass is preserved.  Of course,
only the inner ($r<c_0r_s$) region is actually simulated and the outer part of
the halo remains virtual. We find that the finite dynamical range has only a
minor (much less than induced by the DF) effect on the evolution of the NFW 
cusp. This effect can be further minimized by the
rescaling the  NFW structure to systems with $c > c_0$, at the expense of
neglecting the outer halo regions. In the present work we  used a
fiducial NFW halo with $c_0=3.33$ and have performed the numerical fitting to
the data by rescaling it to halos with larger $c$.
Numerical experiments support the rescaling procedure and show that, for a
given $N$, the rescaled model undergoes much less central flattening than
unscaled models.

The main shortcoming of our numerical model is the treatment of
galaxies as point-like particles of fixed mass. Thus, we neglect the internal
degrees of freedom associated with individual galaxies and their tidal
stripping and merging. The tidal stripping reduces the mass of individual
clumps and their total mass, weakening the DF,  while the merging increases
it. In addition, $N$ used here would not allow for cluster-galaxies resonance
interactions, those may enhance the flattening of the cuspy profile (Katz
2002). A proper treatment of these two processes is  beyond the scope of this
paper and constitutes a challenging problem to cosmological and galaxy
formation simulations. We also neglect the mass spectrum of individual clumps
which only can affect the timescale.

\section{Results}
\label{sec:results}

Several simulations have been performed, spanning NFW parameters and clump
masses. The parameters of the simulations are set within the framework of the
$\Lambda$CDM cosmogony with the WMAP inferred cosmological parameters (Spergel
\etal 2003). The simulation units are scaled in a way that, at 13.5
inner dynamical times,  the DM  mass distribution is best matched by  STE's
optimal  model within $r<50\hkpc$, where their fit was performed.
It thus applies specifically to cluster MS2137-23 with a large central cD
galaxy. The initial conditions of that simulation are scaled to $r_s=164\hkpc$
and $c=5.45$, corresponding to $R_{vir}= 894 h^{-1}$kpc and $M_{vir}= 2.3
\times 10^{14} h^{-1} M_\odot$. The cluster central ($<r_s$) dynamical time
is $\tau_D=0.31 h^{-1}$ Gyr. The softening length, which can be considered as a
characteristic scale length for the clumps, is  $4.2 \hkpc$.

\begin{deluxetable}{lcccccc}
\tablecaption{Model Parameters}
\tablehead{
Model &  $M_{clump}/ ( 10^{11}\hmsun)  $ & $N_{clump} $ &  Mass Fraction in
Clumps & }
\startdata
{\bf 1}  & $1.04 $              & $90$    & $6\%$                  \nl
{\bf 2}  & $1.04 $              & $45$    & $3\%$                  \nl
{\bf 3}  & $1.04 $              & $22$    & $1.5\%$               \nl
{\bf 4}  & $0.35 $              & $900$  & $20\%$               \nl
{\bf 5}  & $0       $              & $0 $     & $0\%$                  \nl
\enddata
\label{table:models}
\end{deluxetable}

Table 1 specifies the parameters of the five different simulations performed
here. Three of the models (1 -- 3) have the same $M_{clump}$, but different
number of clumps. The mass of the clumps in model 4 is a third of that in
model 1 but the number of clumps is 10 times larger. Model 5 is a control run
(DM only, no clumps). Model 1 is considered to be the fiducial nodel, in which
$6\%$ of the total mass is in the clumps. Models 3 and 4 are considered the
extreme cases.

\begin{figure}[ht!!!!!!]
\vbox to2.2in{\rule{0pt}{2.2in}}
\includegraphics{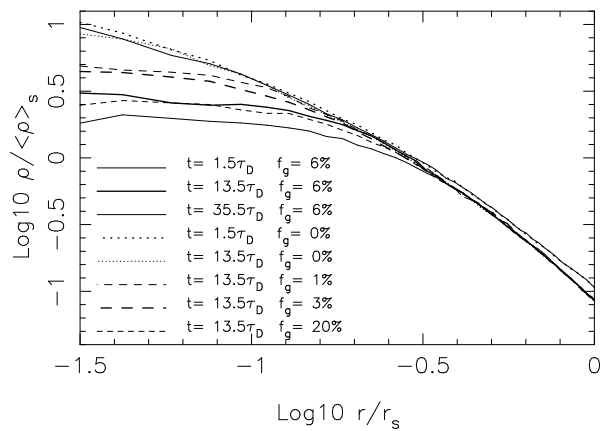}
\caption{The DM density profile of the fiducial model ($f_g=6\%$) is 
plotted at three different times, $t=1.5, 13.5$, and $35.5\tau_D$. The density
is scaled by the average density inside the initial NFW scale length $r_s$, 
$\langle\rho\rangle_s$. The DM density profile of the other models
(control run and the ones with $(M_{clump}/10^{11}\hmsun,f_g)$ = (1.04, 3\%),
(1.04,1.5\%) and (0.35,20\%) is plotted at $t=13.5\tau_D$. The initial density
profile of the unperturbed control model is given for reference.
\label{fig:DM-density}
}
\end{figure}

Fig.~\ref{fig:DM-density} shows the DM density profile of the fiducial model
(1) at $1.5, 13.5$ and $33.5\tau_D$, corresponding to $0.46, 4.2$, and $10
h^{-1}$ Gyr, with the density cusp dissolving with time (the results are shown
in terms of the initial NFW scale length and the mass enclosed within it). In
addition the densities of the other models (2 -- 4; evaluated at
$13.5\tau_D$) and the control case (model 5) of no clumps are shown. At
$13.5\tau_D$, all models with clumps show very clear and pronounced
flattening,  with models 1 and 4 exhibiting very similar profiles. Models 1 --
3 which have the
same $M_{clump}$ show a clear increase of the radius at which the density
profile deviates from its original NFW profile with the total mass in clumps.
The control model has hardly evolved at  $13.5\tau_D$, but shows a small
flattening at $t=35.5\tau_D$, much smaller than the one induced by the massive
clumps. The flattening of the unperturbed NFW cusp can be of dynamical nature
as was noticed before by Kazantzidis, Magorrian \& Moore (2004; for
isotropic velocity dispersion) and/or due to two-body relaxation effect
(Diemand \etal\ 2003). No significant evolution is found outside $r_s$ for all
the models.

Models 1 and 4 differ by a factor of three in $M_{clump}$
and ten in $N_{clump}$, yet they exhibit a very similar DM density profile (at
$t=13.5\tau_D$). This degeneracy follows from the simple Chandrasekhar
formalism. The DF acceleration scales with $M_{clump}$, hence the force is
propotional to $M{_{clump}^2}$. The work done by a single clump on the DM halo
per unit time scales also with $M{_{clump}^2}$. Thus, increasing the mass of
each clump by a factor $3$ and  decreasing their number by a factor $9$, to
get model 4 from model 1, should reproduce the same  density profile.

\begin{figure}[ht!!!!!!]
\vbox to2.4in{\rule{0pt}{2.4in}}
\includegraphics{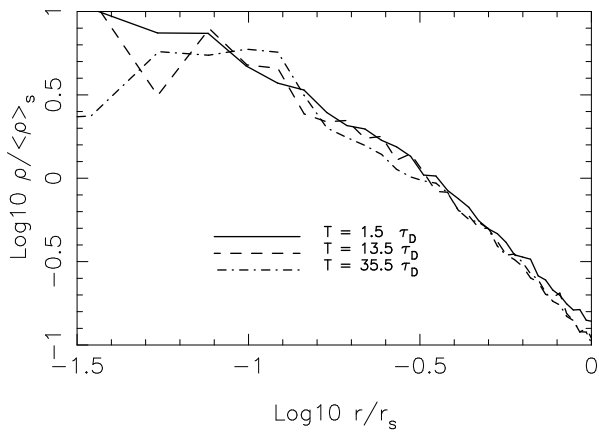}
\caption{The total density profile of the fiducial model ($f_g=6\%$) is plotted
at different times. (Same notations and times as in Fig.\ref{fig:DM-density}.)
\label{fig:TM-density}
}
\end{figure}

\begin{figure}[ht!!!!!!]
\vbox to2.4in{\rule{0pt}{2.4in}}
\includegraphics{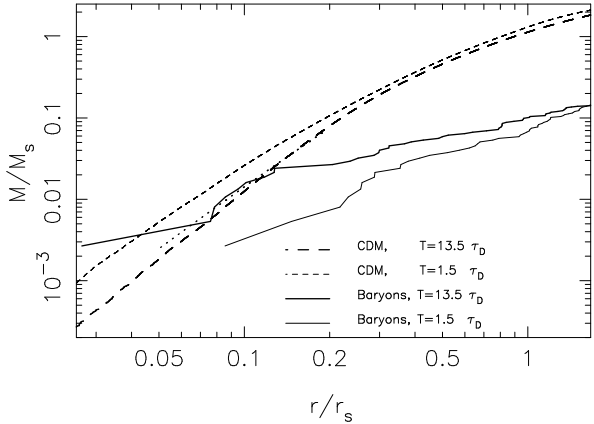}
\caption{The DM and baryon mass profiles (of the fiducial
model, $f_g=6\%$,
in terms of the mass enclosed inside the initial NFW scale length $r_s$)
are plotted at different
times, $t=1.5$ and $13.5\tau_D$. The dotted
curve corresponds to STE's best fit of $\beta=0.35$.
\label{fig:mass}
}
\end{figure}

The total density profile hardly changes at these times (see
the fiducial model in Fig. \ref{fig:TM-density}).
The comparison of Figs. \ref{fig:DM-density} and \ref{fig:TM-density}
implies that the DM cusp dissolves
almost completely, while the clump distribution becomes even more cuspy.
The figures show that density profiles change very little after about
half the age of the universe, pointing to saturation effect (due to the
decreasing density and increasing  local dynamical time and velocities, the DF
time scales roughly as $\sim r^{2.5}$ in an NFW cusp).
They imply also that flattening of the cusp by DF is a robust effect and
no fine tuning of the parameters is necessary, as is evident from
Fig.~\ref{fig:mass}, showing the evolution of
the DM and baryonic mass profiles in this cluster where the baryons dominate
in the center at the present epoch.

STE fit the baryonic and DM density profiles only within the very inner
part of the cluster, roughly within $r=50\hkpc$. STE assumed that the DM
follows a generalized NFW profile,
\begin{equation}
\label{eq:gNFW }
  \rho  = \frac{\rho_c \delta_c} {(r/r_{sc})^{\beta}
   [1+(r/r_{sc})]^{3-\beta}},
\end{equation}
where $\rho_c$ is the critical density,
$\delta$, $r_{sc}$, $\beta$ --- parameters to be fitted to the
data.  Good fits to the evolved CDM mass distributions of our model
can be found in terms of Burkert (1995) profiles, or variations
thereof (ESH). At $13.5\tau_D$, for example, the halo up to
the initial virial radius can be fitted by a Burkert profile with an
RMS error of $\sim 2\%$. The radius of the nearly flat
core is about a third of the initial NFW scale length characterizing
the initial distribution.  The generalized NFW
profiles used by STE cannot fit the final distributions for any value
of the inner $\rho$ slope. Not even $\beta=0$
provides a reasonable fit, since the slope it predicts does not
converge to the original NFW slope outside the core. Therefore
the STE fit implies only a very shallow inner cusp, almost a core, and
cannot be used to infer the structure of the whole cluster. Indeed,
our evolved DM profile follows closely STE's fitted profile,
with $\beta =0.35$ and $\delta=24000$ over
the range of $8 \la r \la 50 \hkpc$, namely  from roughly  twice the
softening length to the maximal radius probed by STE.

\section{Discussion}
\label{sec:disc}

The model considered here, in which the baryons in the cluster are
represented by massive smoothed clumps,
should capture the key effects of DF in the cluster center.
Several effects not explicitly included could further soften the central cusp.
For one thing, the DM halos of individual galaxies, modeled here by the
baryonic clumps, are not included.  These halos will be mostly
tidally stripped before reaching the cluster center, but still
will increase  the speed of DF infall.  Also, galaxies
just falling into the cluster will mostly have rather radial orbits
(Vitvitska \etal 2002), and often fall in as part of groups;
both effects will also increase the rate of infall.
Finally, the (unresolved) dynamical resonances can enhance the cusp softening.

The key prediction of our model is threefold: the total mass distribution is
given by a cuspy NFW-like model, the inner DM density profile is characterized
by a near core structure, and the galaxy distribution becomes steeper than the
original NFW cusp. The first point is in agreement with
observational determinations of the total mass distribution in clusters derived
from X-ray observations by Chandra and XMM (Buote 2002, 2003; Lewis \etal
2003), weak lensing (Dahle, Hannestad \& Sommer-Larsen 2002) and strong
lensing (Bartelmann 2002) --- all consistent with $\beta=1$ or steeper cusps.
The STE analysis of MS2137-23, on the other hand, which separates the DM
distribution from the total matter, excludes an NFW cusp for the DM
distribution and shows consistency with a $\beta=0.35$ --- nearly a flat core.
The predicted steepening of the galaxy distribution is consistent with the
findings of Brunzendorf \& Meusinger (1999), who did an extensive galaxy survey
of the Perseus cluster and found the projected galaxy distribution to diverge
as $r^{-1}$.

The implication of the present paper and of ESH and El-Zant \etal (2004) is
that the cusps in DM halos with clumpy baryons can be washed out by DF.
Most importantly, while the DM and baryon density profiles change dramatically,
{\it the total mass distribution is hardly affected.} The stability of the
total matter distribution persists at least to the present epoch, and is not a
transient phenomenon.
The efficiency of the whole process depends on the nature of the clumpiness:
the mass in the clumps, the clump mass spectrum, and their ability to survive
negative feedback and tidal stripping. It thus depends on the amount of
baryons reaching the center by this process -- clusters without large cD
galaxies are expected to have less flattening of the DM cusp.

We note that the current observational constraints on the
inner halo density profile  may, in principle, be significantly diluted if
the possibility of a nonspherical inner mass distribution is taken
into account (Bartelmann \&  Meneghetti 2003; Dalal \& Keeton 2003).  Our
results should, therefore, be seen as predictions of future observations as
much as explanations for current ones.

Clusters of galaxies provide the `cleanest' case for the operation of
DF, as the clumps (individual galaxies) survive to the present epoch,
and cD galaxies with multiple nuclei
(representing such galaxies in the cluster center) are well
known. Any CDM that is not stripped from the galaxies would contribute
to the final mass distribution parametrized  in STE's analysis
by a Jaffe model, and not to the generalized NFW. It thus enters
implicitely into the stellar M/L ratio inferred from their fits ---
consistent with the local value after accounting for passive evolution.
Thus the ``clumps'' mediating the DF are indeed baryon dominated.
If forming galaxies evolve through clumpy baryonic phase, then the DM density
cusp is expected to be flattened as well (ESH; El-Zant \etal 2004). The
apparent disagreement between dissipationless simulations and
observations of the central density profiles of galaxies and clusters
can be resolved within the CDM cosmogony.

%%%%%%%
\acknowledgments

We thank A. Dekel, R. Ellis, R. Jiminez, N. Katz, A. Klypin, P. A.
Kravtsov, P. Madau, D. Sand \& T. Treu for comments and discussions.
This research has been partially supported by the ISF 143/02 and the Sheinborn
Foundation (to YH), NASA grants NAG 5-12326 (at UCSC), NAG 5-10823, NAG
5-13063 and HST AR-09546.01-A (to IS), and  NSF grants AST-0205944 (at UCSC)
and AST-0206251 (to IS).  IS is grateful to the Lady Davis Foundation for
support. Simulations have been realized on the Fujitsu
NEC-SX5 of the IDRIS-CNRS computing center, at Orsay, France.

\end{document}